\documentclass[a4paper,12pt]{article}
\usepackage[utf8]{inputenc}
\usepackage{amsmath}
\usepackage{breqn}
\usepackage{longtable}
\usepackage{array}
\usepackage{graphicx}
\usepackage[affil-it]{authblk}
\usepackage{csquotes}
\usepackage[
backend=biber,
style=apa,
uniquename = false]
{biblatex}
\DeclareLanguageMapping{english}{english-apa}

\usepackage{textcomp}
\usepackage{gensymb}
\usepackage{multicol,caption}
\usepackage[a4paper, top = 3cm, left = 2cm, bottom = 3.6cm, right = 1.5cm]{geometry}
\graphicspath{ {./images/} }
\newenvironment{Figure}
  {\par\medskip\noindent\minipage{\linewidth}}
  {\endminipage\par\medskip}
\usepackage{supertabular}  
\usepackage{tabularx}

\usepackage{titlesec}
\titleformat*{\section}{\normalsize\bfseries}
\titleformat*{\subsection}{\normalsize\bfseries}
\titleformat*{\subsubsection}{\normalsize\bfseries}
\titleformat*{\paragraph}{\normalsize\bfseries}
\titleformat*{\subparagraph}{\normalsize\bfseries}

\title{The Role of Time, Weather and Google Trends in Understanding and Predicting Web Survey Response}
\author[1]{Qixiang Fang}
\author[2]{Joep Burger}
\author[2]{Ralph Meijers}
\author[2]{Kees van Berkel}
\affil[1]{Utrecht University}
\affil[2]{Statistics Netherlands}
\date{\vspace{-5ex}}

\begin{document}
\maketitle

\begin{abstract}
In the literature about web survey methodology, significant efforts have been made to understand the role of time-invariant factors (e.g. gender, education and marital status) in (non-)response mechanisms. Time-invariant factors alone, however, cannot account for most variations in (non-)responses, especially fluctuations of response rates over time. This observation inspires us to investigate the counterpart of time-invariant factors, namely time-varying factors and the potential role they play in web survey (non-)response. Specifically, we study the effects of time, weather and societal trends (derived from Google Trends data) on the daily (non-)response patterns of the 2016 and 2017 Dutch Health Surveys. Using discrete-time survival analysis, we find, among others, that weekends, holidays, pleasant weather, disease outbreaks and terrorism salience are associated with fewer responses. Furthermore, we show that using these variables alone achieves satisfactory prediction accuracy of both daily and cumulative response rates when the trained model is applied to future unseen data. This approach has the further benefit of requiring only non-personal contextual information and thus involving no privacy issues. We discuss the implications of the study for survey research and data collection.

\vspace{2mm}
\end{abstract}

\vspace{2mm}
\begin{multicols}{2}
\section{Introduction}
Web surveys have become increasingly popular over the past decades. This growing popularity of web surveys, in comparison with their traditional counterparts (e.g. telephone and face-to-face interviews), can be attributed to unique advantages of web surveys such as shorter transmission time, lower costs and more flexible designs \parencite{JelkeBethlehem1}. However, web surveys also suffer from various issues, the most prominent of which are lower response rates, which leads to compromised quality of the resulting survey data (i.e. sometimes more bias and always less precision) (e.g. \cite{Taylor2019}; \cite{Fowler2019}; \cite{Fan2010}; \cite{Manfreda2008}). A potential solution are mixed-mode designs, which follow up on web surveys with other modes like telephone and face-to-face interviews \parencite{Wolf2016}. In this way, the total response rate becomes higher, the collected responses more representative and the resulting estimates less biased. Nevertheless, low response rates of the web mode in mixed-design surveys are still undesirable because of the associated increase in total expenses and planning effort. Therefore, in many ways, low response rates are a challenge for web surveys, necessitating research to understand the likely mechanism underlying unit (non-)response decisions in web surveys \parencite{Fang2012a}. 

The past decades has seen a plethora of research on influencing factors of web survey (non-)response. Most of these factors belong to the following three categories: respondent-related (e.g. age, income, education), region-related (e.g. degree of urbanisation, population density) and design-related (e.g. survey length, contact mode). For a quick (non-exhaustive) overview of existing findings in the web survey literature, see Appendix A. These discoveries have greatly improved the understanding of the underlying mechanisms of response decisions in web surveys. However, they do not fully account for the variation in observed response behaviour. For instance, \textcite{Erdman2016} used the best twenty-five out of over three hundreds of such predictors in their study to predict block-level response rates and yet the resulting model only explains about 56\% of the variation in the response rates. It remains, therefore, necessary to investigate additional influencing factors. 

One prominent characteristic of the three categories of factors described above is that they are \textit{time-invariant}, meaning that they tend to stay constant during a survey's data collection period. Even when they do vary in values, the change is unlikely substantial. The time-invariance of these well-studied factors (partially) explains why the use of them alone is insufficient to substantially explain variations in (non-)response patterns, simply because they cannot properly account for temporal fluctuations of survey response rates over time like different days of the week (e.g. \cite{Faught2004}; \cite{Sauermann2013a}), months of the year (e.g. \cite{Losch2002}; \cite{Svensson2012}) and different years \parencite{BartelSheehan2001}. This leads us to look into the counterpart of time-invariant factors, namely time-varying factors such as \textit{day of a week}, \textit{weather}, \textit{presence of public holidays}, \textit{disease outbreaks} and \textit{salience of societal issues} like privacy concerns, to name a few. 

The lack of research on the roles of such time-varying factors not only limits our understanding of the underlying process of response decisions in web surveys, but also hinders survey design effort aiming at increasing response rates. For instance, people may be unlikely to respond to surveys in holiday periods because they are not at home and/or do not want to spend time on surveys. Therefore, survey response rates might be higher or lower depending on time-related factors. This is also relevant to surveys that accept responses over a longer period of time. In such surveys, daily response rates usually peak on the first day(s) and quickly subdue \parencite{Minato15}. This means that the number of responses during the first day(s) significantly influence the final response rates of the surveys, suggesting that effects of time-varying factors (if there is any) during the first few days might be crucial for a proper understanding of the final response rates. 

Therefore, in this paper, we investigate whether and how various time-varying factors (including time, weather and societal trends) influence the daily (non-)response patterns of the web mode of the 2016 and 2017 Dutch Health Surveys, and whether the use of these factors are can be useful in predicting survey response rates. 

We structure the remainder of the article as follows. We begin with a detailed account of time-varying (contextual) factors and our research proposal, followed by the research aims. Then, we describe the data, research methods and results, respectively. We conclude with a discussion on the theoretical and practical implications of the findings and recommendations for future research. 

\section{Time-Varying Contextual Factors}
\subsection{Definitions}
In contrast to time-invariant factors, time-varying factors are variables whose values may differ over time \parencite{Singer2003}. They record an observation’s potentially differing status on each associated measurement occasion. Some have values that change naturally; others have values that change by design. For survey research, which has a typical fieldwork period ranging from a few days to several months, it is likely that some underlying factors of (non-)response behaviour vary substantially in their values during the fieldwork period. Examples are: personal availability, individual emotional status, day of a week, weather, holidays, and societal sentiments about a topic. Of course, respondent- and region-related factors can also vary (e.g. gender, age, household composition, geographical features of the city of residence), but normally to a much lesser extent (especially during a typical survey project period) and unlikely apply to the majority of the sample units. Therefore, we argue that time-varying factors can help to understand and predict survey responses, especially with regards to fluctuations in response rates over time which time-invariant factors by definition cannot properly account for. 

In this paper, we focus on time-varying factors that are also contextual factors. We define contextual factors as variables that usually cannot be influenced by study participants because they are determined by stochastic processes largely or totally external to them. Typical contextual factors assess potentially changing characteristics of the physical or social environment in which study participants live. Some of the aforementioned examples of time-varying factors like weather, time and societal sentiments fit this definition. The term, like others, is defined and used differently across disciplines. In survival analysis research, it is often called ``ancillary factors" \parencite{Kalbfleisch2002}. In this paper, we prefer to use the more intuitive term ``contextual factors".

The focus on only contextual factors has two practical purposes. The data for contextual factors are usually non-personal, meaning that they do not involve privacy issues. This is important because recent increases in general privacy concerns among the public and the introduction of more stringent privacy regulations such as the General Data Protection Regulation in the EU have led to greater difficulties in obtaining, accessing and using personal data. Our attempt to model web survey response with only non-personal predictors, if proven successful, can be a useful alternative for the survey research community. Furthermore, because contextual factors are (largely) free from the influence of study subjects, there is no or little concern for reverse causality and hence more internal validity for the study.

To sum up, in this paper, we focus on predicting and understanding web survey responses from time-varying contextual factors. 
\subsection{Literature Review}
A significant amount of research in fields like cross-cultural psychology, epidemiology and family sociology has investigated the effects of time-varying contextual factors on individual outcomes. In contrast, much fewer studies in the field of survey research have attempted so and even fewer focus on web surveys. Therefore, for a more thorough understanding of the likely effects of time-varying contextual factors, we conduct a literature review on this topic including not only web survey studies but also non-web survey ones.

The studies on time-varying contextual factors that we identified can be categorised into two types. The first type focuses on the effect of time, such as year, season, month, days of a week. For instance, \textcite{BartelSheehan2001} analysed 31 web surveys and concluded that the \textit{year} in which a survey was published was the most important predictor of response rates. \textcite{Losch2002} found that completing a survey interview during \textit{summer} in Iowa (US) required more contact attempts than in other seasons. Similarly, \textcite{Goritz2014} documented for a German online panel that the panel members were more likely to start and finish studies in \textit{winter} than during any other season. Contrasting these two findings, \textcite{Svensson2012} in a Swedish longitudinal online survey found that the highest response rate was in \textit{September}. \textcite{Faught2004} noted in their experimental study on US manufacturers that survey response rates were the highest when the email invitation was sent on either \textit{Wednesday morning} or \textit{Tuesday Afternoon}. Contrary to this, \textcite{Sauermann2013a} in their experiments, which was conducted among US researchers, did not find the timing of the e-mail invitation (in terms of the day of a week) to result in significantly different response rates in a web survey. However, they did find that people were less likely to respond in the \textit{weekend} and would postpone the response until the next week. 

The second type of studies concerns the influence of weather on survey participation. \textcite{Potoski2015} analysed eight surveys from 2001 to 2007 and showed that on unusually \textit{cold} and \textit{warm} days, wealthier people are more likely to participate in surveys than the less wealthy. \textcite{CunnighamR1979} found more \textit{pleasant weather} (e.g. more sunshine, higher temperature, lower humidity) to significantly improve a person's willingness to assist an interviewer. 

The effect of \textit{weather} on survey participation, however, likely goes beyond these two findings. \textcite{Simonsohn2010} showed that on \textit{cloudier} days people are more likely to engage in academic activities, which share some common characteristics with survey participation (e.g. high cognitive load and low immediate returns). Therefore, it is likely that people on cloudy days may become more inclined to participate in surveys. \textcite{Keller2005} showed that higher air pressure has a positive influence on mood. In turn, positive mood states may lead to increased helping behaviour (e.g. fulfilling a survey request) \parencite{Weyant1978}. Therefore, air pressure may also impact survey response decisions. 

\subsection{Research Proposal}
The studies above do confirm, to some degree, the influence of time-varying contextual factors on response decisions in surveys. However, some contradictory findings suggest a need for further research, for instance, on the influence of \textit{day of a week}. In addition, it is also likely that the results of the existing studies are confined to local applications and that findings based on non-web surveys may not apply to web surveys. New time-varying contextual factors should also be researched. We propose the following time-varying contextual factors for our study. 

\subsubsection{Time}
The first category of time-varying contextual factors is related to time, including \textit{day of a week} and \textit{public holidays}. The former is chosen because according to the literature, its effect on survey response rates is still unclear, seemingly varying across surveys and thereby needing further research. The latter factor is chosen because we hypothesise that during holidays, people may travel around, spend more time with family or want to rest and are consequently less likely to participate in survey research. 

\subsubsection{Weather}
The effect of weather on survey participation also requires more research. In this study, we include different types of daily weather measures (e.g. maximum, minimum and average) of temperature, sunshine, precipitation, wind, cloud, visibility, humidity and air pressure. 

\subsubsection{Societal Trends}
In addition to time and weather, factors which relate to real-time societal trends such as disease outbreaks, privacy concerns, public outdoor engagement (e.g. in festivals and on the road), terrorism salience may also play an influencing role in affecting survey participation decisions. These we term ``\textit{societal trends}" in this paper. We explain each of these societal trends factors next.

\paragraph{Disease Outbreaks}
It is common knowledge that being sick (physically or psychologically) can alter individual behaviour. For instance, individuals who are sick may stay at home more and reduce outdoor or professional activities. They may lack the cognitive resources to engage in cognitively demanding activities (like survey participation). They may develop more negative emotions, which in turn may influence their pro-social behaviour. In particular, medical conditions such as the common cold, the flu, hay fever and depression are likely to affect a large number of individuals, especially during certain times of a year (e.g. cold, flu and depression in the winter; hay fever in the spring). Therefore, we hypothesise that disease outbreaks can have physical, behavioural and psychological consequences that in turn impact survey participation to a varying degree depending on the type, severity and prevalence of the illness. We did not to find any previous research on this topic, leading us to believe that our hypothesis is novel and worth studying.

\paragraph{Privacy Concerns}
Research has shown that privacy concerns can deter individuals from survey participation. For instance, two studies on the 1990 and 2000 U.S. census find that an increase in concern about privacy and confidentiality issues is consistently associated with a decrease in the probability of census participation, especially among certain ethnic groups (\cite{singer1993}; \cite{SingerHoewyk2003}). Using paradata, two other studies report that greater privacy concerns are linked to higher unit or item non-response (\cite{Dahlhamer2008}; \cite{Bates2008}). Given these findings, we hypothesise that the level of general societal concerns about data privacy issues likely predicts survey response rates. Specifically, the higher the levels of the privacy concerns, the lower the response rates.

\paragraph{Public Outdoor Engagement}
Considering the nature of some types of surveys (e.g. mailed surveys, digital surveys that require a desktop and are not smartphone-friendly), it is conceivable that people are unlikely (if not impossible) to participate in surveys when they are engaged in outdoor activities (e.g. public events, holidays, travel), even if they have received the survey requests. This is evidenced by the previous finding that completing a survey interview during summer required more contact attempts than in other seasons because people are more likely travelling \parencite{Losch2002}. Therefore, we hypothesise that the level of outdoor engagements in during certain time periods may also affect survey response rates during that time. This can be especially true for the surveys that we study in this research, where survey invitations are mailed to the sample units’ home addresses and the surveys are not suitable for smartphones (i.e. they require the use of laptops or tablets). 

\paragraph{Terrorism Salience}
A large body of literature has shown that terrorist events have impactful individual and societal consequences. For instance, higher levels of terrorism salience or fears are linked to more negative emotions for non-religious people \parencite{Fischer2006}, worse mental health \parencite{Fischer2008}, more media consumption (\cite{Boyle2004}; \cite{Lachlan2009}), increased contact with family and friends \parencite{Goodwin2005}, (irrational) travel behaviour (\cite{Gerd2006}; \cite{Baumert2019}), temporarily lower social trust \parencite{Geys2017} but more institutional trust (\cite{Sinclair2010}; \cite{Dinesen2013}), less occupational networking activities \parencite{Kastenmueller2011}, cancellation of sport events and a higher number of no shows in sport events \parencite{Frevel2020}. Furthermore, mortality salience (which can be induced by reports of deaths in terrorist events) has been shown to increase pro-social attitudes and behaviour \parencite{Jonas2002}. These studies were conducted across various countries, related to different terrorist events and with study participants either directly or indirectly impacted by terrorist events. These consistent findings lead us to reason that terrorism salience or fears can have a substantial impact on most members of a society, including potential survey respondents. In the specific context of survey research, the findings that terrorism salience or fears can induce emotional and behavioural changes in, for example, health status, media use, travel behaviour and interpersonal relationships suggest that survey participation (and consequently, survey response rates) can also be indirectly affected. Given all the potential consequences of higher terrorism salience (which likely take greater priorities in one’s life than survey participation), we tentatively hypothesise a negative effect of terrorism salience on survey response rates. Furthermore, note that the Netherlands (where the current study is based) and its nearby countries (such as Germany and France) have suffered from terrorist threats, attacks or related events and issued terrorism warnings during the past years (see \cite{Terrorism2016}; \cite{Terrorism2017}). This fact makes our inclusion of terrorism salience as a potential factor of survey response all the more relevant.

\subsubsection{Summary}
Some of the existing studies investigated effects of time-varying factors on a monthly or yearly scale. These findings are certainly interesting and informative; nevertheless, studying the effects of time-varying factors on a finer scale (e.g. weekly, daily) might provide survey researchers with even more helpful insights. In this project, we focus on the effects of \textbf{daily} time-varying factors on daily survey response. We do not consider the effects of months or seasons here, partly because we expect the daily time-varying variables we use to be able to capture any monthly and seasonal trends and partly because there is not sufficient variation in our data to allow for reliable estimation of the relevant month and season effects that would generalise well to future unseen data. 

To sum up, we propose the following daily time-varying factors for investigation in the current study: \textit{day of a week}, \textit{public holidays}, \textit{weather (i.e. temperature, sunshine, precipitation, wind, cloud, visibility, humidity and air pressure)} and \textit{societal trends} (i.e. \textit{disease outbreaks}, \textit{data privacy concerns}, \textit{public outdoor engagement} and \textit{terrorism salience}). 

\section{Research Aims}
The current study is mainly concerned with whether and how daily time-varying contextual factors such as day of a week, holidays, weather and societal trends influence web survey response behaviour on a daily basis. To approach this question, we use discrete-time survival analysis to model the effects of predictors on the daily conditional odds of a person responding to a web survey (given that he/she has not responded yet). In this way, we obtain insight into how a specific factor influences the daily response decision of a person. 

Furthermore, believing that a model (or a predictor) is generally more useful when it not only explains the current data but also generalises to future observations, we evaluate the trained models and the related predictors with regards to their predictive performances on an independent data set. This approach helps to answer, for instance, whether \textit{temperature} is a better predictor than \textit{day of a week}.

\section{Data}
\subsection{The Dutch Health Surveys}
In this study, we analyse the response decision of individuals who were invited to participate in either the 2016 or the 2017 Dutch Health Surveys. The Dutch Health Survey is a yearly survey administered by Statistics Netherlands. It aims to provide an overview of the developments in health, medical contacts, lifestyle and preventive behaviour of the Dutch population. The sampling frame comprises of persons of all ages residing in private households. It utilises a mixed-mode design, consisting of an initial web mode and follow-up telephone or face-to-face interviews in case of non-response in the web mode. Only the design and the data of the web mode are relevant to this study. 

\begin{Figure}
\includegraphics[width=\linewidth]{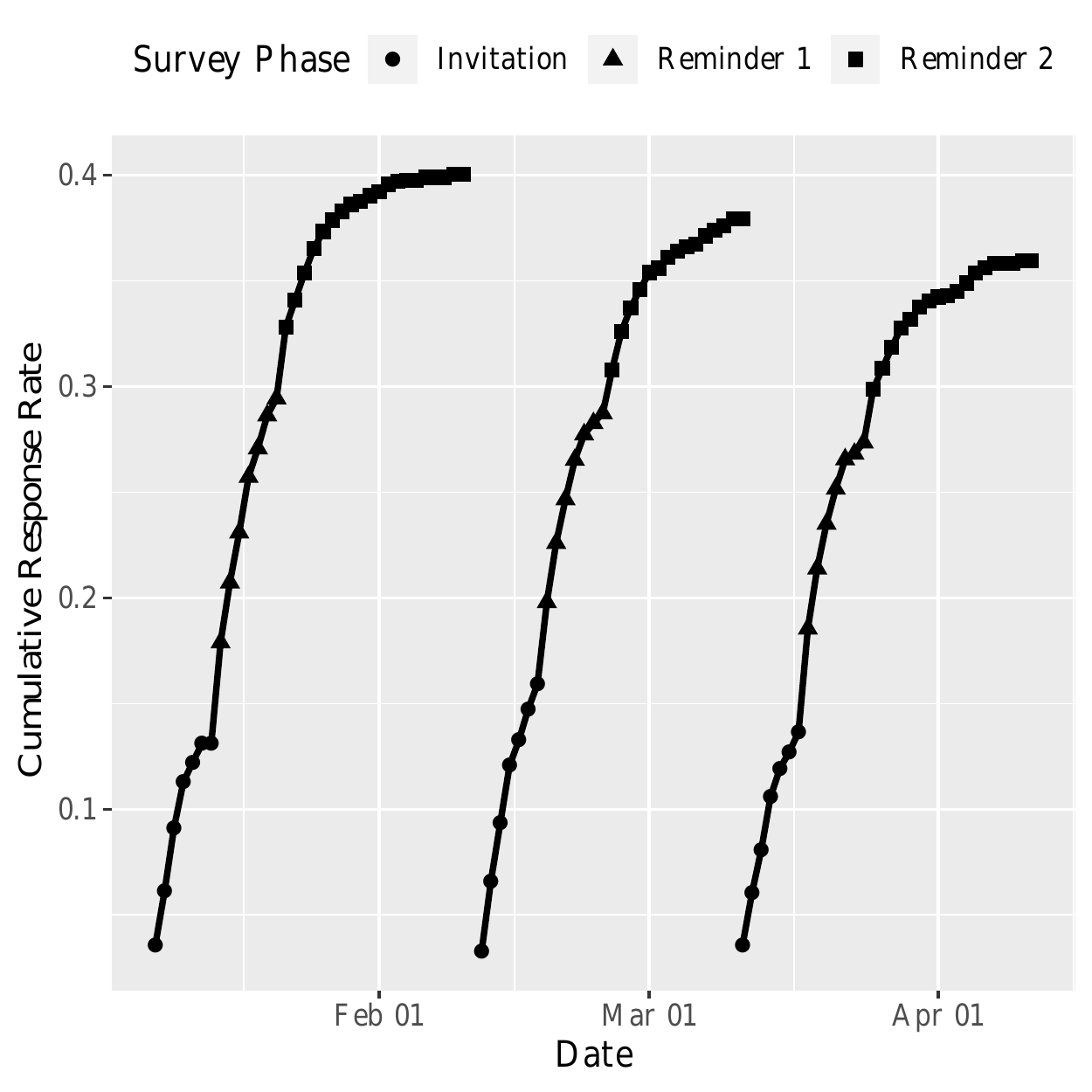}
\centering
\captionof{figure}{Cumulative Response Rates of Three Web DCPs in the 2016 Dutch Health Survey}
\end{Figure}

A yearly sample is divided into 12 cohorts, each corresponding to a data collection period (DCP). Each DCP starts with a web-mode survey, which lasts about a month. A web-mode DCP begins with the \textbf{by-post} delivery of an invitation letter that requests the sample unit to respond to the survey online using a desktop, laptop or tablet (but not a smartphone). In case of no response from the individual after about one week, up to two mailed reminder letters follow (at an interval of one week). The invitation and reminder letters contain a web link to the survey and a unique personalised password required for login to the survey. Each web-mode DCP ends roughly one month following the invitation letters. 

Figure 1 illustrates the data collection process with the cumulative response rates of the first three web-mode DCPs of the 2016 Dutch Health Survey. The starting point of a new curve indicates receiving a corresponding invitation letter. Each web-mode DCP ends at the end of the curve. We can see that response rates grow the fastest in the first few days after an invitation or reminder letter and flatten quickly later, suggesting that the first few days of data collection are crucial for ensuring a high response rate. 

\begin{minipage}[t]{.45\textwidth}
\captionof{table}{Comparison of the 2016 and 2017 Dutch Health Survey (Web Mode)}
\begin{tabularx}{\textwidth}{p{2.9cm}p{2.9cm}p{1.7cm}}
\hline
 & \textbf{2016} & \textbf{2017} \\ \hline
Expected Delivery Day of Invitation & Fri. (Jan.-Jun.) Sat. (Jul.-Dec.) & Thur. \\\hline
Expected Delivery Day of Reminder & Fri. (Jan.-Jun.) Sat.   (Jul.-Dec.) & Sat. \\
\hline
Sample Size & 15007 & 16972 \\
\hline
Response Rate & 34.8\% & 34.2\% \\
\hline
\end{tabularx}
\end{minipage}

\vspace{0.5cm}
The Dutch Health Survey has a relatively consistent survey design over the years (up to and including 2017), thereby making comparison and integration of data from different years valid and simple. Table 1 summarises information about expected delivery days of the letters, sample sizes and response rates of the 2016 and 2017 surveys (web-mode). Note that in the first half of 2016, the letters were scheduled to arrive on Friday, while in the latter half on Saturday. In contrast, the invitation and reminder letters in 2017 were expected to arrive on Thursday and Saturday, respectively. This variation in the expected arrival days of the letters (i.e. variation in the designs of the surveys) can increase the robustness of the study results, especially with regards to the effects of the \textit{day of a week} predictor. 

While we acknowledge that there are different categories of non-response behaviour such as non-contact and refusal (e.g. \cite{Lynn2002}) and that there is research value in differentiating the sub-types of non-response behaviour, we treat all non-response sub-categories as one single ``non-response" category in this study. There are two reasons. First, the focus of our study is on response and non-response. In this sense, we are not interested in the sub-types of non-response behaviour. Second, the number of non-contact and refusals in our data is too low (non-contact rates $<$ 0.4\% and refusal rates $<$ 2.2\%) for the use of discrete-time survival models in this study, because the denominator of hazard rates becomes so low that the models would have trouble with reliable estimation.

\subsection{Weather Data}
The Royal Netherlands Meteorological Institute (KNMI) records daily weather information about temperature, sunshine, precipitation, wind, cloud, visibility, humidity and air pressure across 47 stations in the Netherlands. We retrieved the 2016 and 2017 daily weather information (i.e. 20 variables in total) from the KNMI website \parencite{KNMI2019}. The exact variables and the associated measures are summarised in Appendix B. 

We averaged the obtained weather records across all stations, instead of assigning every sample case to the nearest weather station, for two reasons. First, considering the small size and geographical homogeneity of the Netherlands and only small variations of the weather data across the weather stations, we did not see a strong benefit of assigning the closest weather station records to the sample cases over simply assigning the average scores, especially when we also factor in the additional effort and potential matching mistakes associated with the matching approach. Second, averaging the weather measures across stations means that all sample cases, on a given day, are associated with the exact same scores for any weather variable. This has the advantage that we can collapse the data set from the ``Person-Period" data format into the ``Period-Level" data format (see Section 5.2 for more information), which has the advantage of downsizing the data matrix and thus reducing model computation time (while obtaining the exact same model estimates). Therefore, given both considerations, we decided to average the weather records across stations.

\subsection{Societal/Google Trends}
The four societal trends of interest are \textit{disease outbreaks}, \textit{privacy concerns}, \textit{public outdoor engagement} and \textit{terrorism salience}. To our knowledge, there are currently no publicly available administrative or survey data on any of these four societal trends on a daily basis in the Netherlands. Measuring these trends, therefore, requires innovative solutions. Our solution of choice is to use Google Trends (GT) data to capture signs of these societal trends.

GT offers periodical summaries of user search data from 2004 onwards for many regions and for any possible search term. These summaries, available as indices, represent the number of Google searches that include a given search term in a specified period (e.g. day, week or month). The data are scaled for the requested period between 0 and 100, with 0 indicating no search at all and 100 the highest search volume in that period. These indices, which represent the popularity of specific search terms, may offer relevant insights into various human activities in (almost) real time. Indeed, GT indices have been used for various purposes, such as real-time surveillance of disease outbreaks \parencite{Carneiro2009}, economic indicators \parencite{Choi2012} and salience of immigration and terrorism \parencite{Mellon2014}.

These successful applications of GT indices suggest the possibility of using GT to capture our four societal trends of interest. Appendix B lists the search terms we used to measure each trend. For \textit{disease outbreaks}, we used the relevant commonly used Dutch terms concerning diseases like ``flu'', ``cold'' and ``depression''. For \textit{privacy concerns}, we used terms such as ``data leaks'' and ``hacking'' as proxies. For \textit{outdoor engagement}, we used terms indicating whether people are in a traffic jam or participating in festivals. Lastly, for \textit{terrorism salience}, we used the term ``terrorist". Note that we hypothesised these search terms prior to any analysis, rather than cherry-picked from a long list sorted by correspondence with response rates, to reduce the problem of spurious correlations. 

Like any other data, GT data also need to be checked with regards to their validity and reliability before use. A potential issue of validity in GT concerns the fact that the search volume of a specific search term does not necessarily measure the intended phenomenon. A key reason is that a search term can bear multiple meanings. For instance, the Dutch term ``AVG'' can be short for both ``Algemene verordening gegevensbescherming (General Data Protection Regulation)'' and ``AVG Technologies'' (a security software company). According to Google Trends, the second meaning was more often used than the first meaning in both 2016 and 2017 in the Netherlands. Therefore, GT indices may capture noise rather than the intended trends of interest. To mitigate this validity issue, we followed the advice by \textcite{Zhu2012}. The authors noted that GT offers a function to check the most correlated queries (i.e. search terms) and topics for any specific term you enter. The validity of the terms can thus be manually assessed by checking whether the most correlated queries and topics correspond to the intended meaning of the search term. For instance, GT shows that the most correlated topic with the Dutch term ``files'' (meaning ``traffic jam'') was ``traffic congestion" in both 2016 and 2017, therefore indicating the relatively high construct validity of the term ``files". Appendix D summarises all the correlated queries and topics of the used GT search terms in this study. 

In addition, the resulting index scores from GT can vary substantially across different requested periods and inquiry attempts. This speaks of measurement reliability issues, which likely stems from two reasons. First, Google Trends uses a simple random sample from the total search volume to calculate the index scores. Therefore, the resulting index scores are subject to high variability if the (unknown) sample size is small. Second, Google does not publish details about the underlying algorithms that calculate the scores, nor does Google make publicly available any changes in its algorithms. Therefore, the obtained index scores may change over time because of updates in the algorithms, even when the exact same search strategy is used. To overcome this issue, we used repeated sampling to enhance the measurement reliability of GT data. By taking as many samples as possible per date in the period of interest and averaging all the scores for each date over all repeated samples, one can obtain much more precise estimates. However, in this procedure a complication arises due to the fact that a different sample can only be taken when a different period is requested. For instance, to obtain two different repeated samples for the date ``2017-01-05", one needs to request two different periods such as ``2017-01-04 to 2017-01-05" and ``2017-01-05 to 2017-01-06", both covering the date of interest (``2017-01-05"). However, as GT scales the scores within the specified period to be between 0 and 100, when two different periods are requested, the reference values in the two groups can be different, leading to differently scaled scores and invalid comparisons of scores between the two different periods. We overcame this issue by maximising the number of overlapped dates between every two consecutive different periods and then calibrating the latter period to the previous one. In this way, the comparability between different samples is maximised. See Appendix E for a detailed description of the algorithm we created and used to calibrate GT data.

\subsection{Overview of Variables}
Appendix B summarises all the variables used in the study. Unless specified as categorical or dummy-coded, the variables are treated as continuous. 

\section{Methods}
In this section we detail the analytic approaches used in the study. First, we introduce discrete-time survival analysis and under this analytical framework, demonstrate how we used logistic regression to model the effects of time-varying factors on survey response in the web mode of the 2016 and 2017 Dutch Health Surveys. We explain how we applied (adaptive) Lasso regularisation with logistic regression, with the goal to enhance model interpretability and predictive performance.

Following the modelling approaches mentioned above, we trained three models with the time-varying contextual predictors based on the \textbf{training data set}, which consists of the complete 2016 Dutch Health Survey data and the first half of the 2017 data (i.e. 18 web-mode DCPs in total). The data is split in this way because it is important to have a good trade-off between enough data variation in the training data and sufficient independence of the test data from the training data. 

Then, we applied the trained models to the \textbf{test data set}, which is the remaining 2017 data (i.e. the last six web-mode DCPs), evaluated and compared their predictive performances. These three models are: the ``baseline model", which includes only the baseline predictors (the number of ``days" since the previous invitation or reminder letter and ``survey phase") that are necessary for the specification of the intercept term in a discrete-time survival model (see Section 5.3); the ``full model", which includes all the time-varying contextual predictors; and, lastly, the ``interaction model" where we allow the effects of the predictors to vary with time. 

Appendix C provides descriptive statistics about all the variables used in the training and the test data sets, separately. 

\subsection{Discrete-Time Survival Analysis}
The research questions and the features of the current data require a modelling framework capable of handling the following issues: first, the method should model the transition from non-response to response; second, it incorporates both time-varying and time-fixed predictors; third, it takes care of the right censoring issue in the data, with right censoring meaning that for some individuals the time when a transition (i.e. response) takes place is not observed during the survey's web mode.

These specific issues call for survival analysis. Survival analysis is a body of methods commonly used to analyse time-to-event data \parencite{Singer2008}. The focus is on the modelling of transitions and the time it takes for a specific event to occur. The current research interests lie in the modelling of the transition from non-response to response over a period of time, thereby making survival analysis the right analysis tool. Many survival analysis techniques (e.g. Cox regression) assume continuous measurement of time. However, in practice, data are often collected in discrete intervals, for instance, days, weeks and months. In this case, a sub-type of survival analysis is needed, namely, discrete-time survival analysis. Given that our data are measured in daily intervals, it is only appropriate to use discrete-time survival analysis.

There are further advantages to using discrete-time analysis, in comparison to its continuous-time counterpart \parencite{Tutz2016}. For example, discrete-time analysis has no problem with ties (i.e. multiple events occurring at the same time point) and it can be embedded into the generalised linear model framework, as is shown next.

\subsection{The General Modelling Approach}
The fundamental quantity used to assess the risk of event occurrence in a discrete-time period is hazard. Denoted by $h_{is}$, discrete-time hazard is the conditional probability that individual $i$ will experience the target event in time period $s$, given that he or she did not experience it prior to time period $s$. This translates into, in the context of this paper, the probability of person $i$ responding to the survey during day $s$ given the individual did not respond earlier. The value of discrete-time hazard in time period $s$ can be estimated as the ratio of the number of individuals who experience the target event in time $s$ to the number of individuals at risk of the event in time $s$. 

A general representation of the hazard function that connects the hazard $h_{is}$ to a linear predictor $\eta$ is

\begin{equation} 
\eta = g(h_{is}) = \gamma_{0s} + x_{is}\gamma
\end{equation}

where $g(.)$ is a link function. It links the hazard and the linear predictor $\eta = \gamma_{0s} + x_{is}\gamma$, which contains the effects of predictors for individual $i$ in time period $s$. The intercept $\gamma_{0s}$ is assumed to vary over time whereas the parameter $\gamma$ is fixed. Since hazards are probabilities restricted to the interval [0, 1], a natural, popular candidate for the response function $g(.)$ is, among others, the logit link. The corresponding hazard function becomes

\begin{equation} 
h_{is} = \exp(\eta)/(1+\exp(\eta)) 
\end{equation}

Under this logistic model, the exponential term of a parameter estimate quantifies the difference in the value of the conditional odds (instead of hazards) per unit difference in the predictor. The total negative log-likelihood of the model, assuming random (right) censoring, is given by

\begin{equation} 
-l \propto \sum_{i=1}^n\sum_{s=1}^{t_{i}} y_{is}\log h_{is} + (1-y_{is})log(1- h_{is})
\end{equation}

where $y_{is} = 1$ if the target event occurs for individual $i$ during time period $s$, and $y_{is} = 0$ otherwise; $n$ refers to the total number of individuals; $t_i$ the observed censored time for individual $i$. This negative log-likelihood is equivalent to that of a binary response model. This analogue allows us to use software designed for binary response models (e.g. binary logistic regression) for model estimation, with only one modification, namely that the number of binary observations in the discrete survival model depends on the observed censoring and lifetimes. Thus, the number of binary observations is $\sum_{i = 1}^{n}\sum_{s = 1}^{t_{i}}$. This requires the so-called Person-Period data format, where there is a separate row for each individual $i$ for each period $s$ (``day'' in our case) when the person is observed. In each row a variable indicates whether an event occurs. The event occurs in the last observed period unless the observation has been censored. Table 2 shows an exemplar Person-Period data set.

\begin{minipage}[t]{.85\linewidth}
\centering
\captionof{table}{Example of Person-Period Data}
\begin{tabularx}{\textwidth}{cccc}
\hline
\textbf{Person} & \textbf{Event} & \textbf{Time} & \textbf{Covariate} \\ 
$i$ & $y_{is}$ & $s$ & $x_{is}$\\ \hline
1 & 0 & 1 & $x_{1,1}$ \\
1 & 0 & 2 & $x_{1,2}$ \\
2 & 0 & 1 & $x_{2,1}$ \\
2 & 0 & 2 & $x_{2,2}$ \\
$\vdots$ & $\vdots$ & $\vdots$ & $\vdots$ \\
2 & 0 & 20 & $x_{2,20}$ \\
3 & 0 & 1 & $x_{3,1}$ \\
$\vdots$ & $\vdots$ & $\vdots$ & $\vdots$ \\
\hline
\end{tabularx}
\end{minipage}

\vspace{2mm}
One may wonder whether the analysis of the multiple records in a Person-Period data set yields appropriate parameter estimates, standard errors and goodness-of-fit statistics when the multiple records for each person in the data set do not appear to be independent from each other. This, fortunately, is not an issue in discrete time survival analysis because the hazard function describes the \textit{conditional} probability of event occurrence, where the conditioning depends on the individual surviving until each specific time period $s$ and his or her values for the substantive predictors in each time period \parencite{Singer2008}. Therefore, records in the Person-Period data need to only assume conditional independence. 

Note that when a data set contains only time-varying predictors and these predictors only vary with time but not with individuals, we can collapse the data set from the ``Person-Period" format into the ``Period-Level" format, where each row represents a given time point, the scores of the time-varying predictors associated with that time point, the number of individuals experiencing the target event and the number of individuals at risk of the event at that time. This approach allows us to significantly downsize the data matrix (from 775,890 rows to only 808 rows, in this study), and thus reduce model computation time (from tens of hours to only minutes, in this study), while obtaining the exact same model estimates as we would with the original Person-Period data format. The only difference is that, instead of using a binary logistic regression, we need to use a binomial logistic regression which models count and proportion outcomes. Our data qualifies for such a transformation and thus we adopt this transformation strategy. 

\subsection{Model Specification}
An important consideration concerns the specification of the intercept $\gamma_{0s}$ shown in Equation 1. $\gamma_{0s}$ can be interpreted as a baseline hazard, which is present for any given set of covariates. The specification of $\gamma_{0s}$ is very flexible, varying from giving all discrete time points their own parameters to specifying a single linear term. In the 2016 and 2017 Dutch Health Surveys, an individual becomes ``at risk" (i.e. of responding to the survey) when he/she receives an invitation letter. Thus, each time point $s$ can be conceptualised as a linear combination of the number of days since the expected delivery of \textit{the previous invitation or reminder letter} (i.e. \textit{days}, a continuous variable), and the specific survey phase this time point is in (\textit{Survey Phase}, a categorical variable with levels ``Invitation", ``Reminder1" and ``Reminder 2"). 

Note that we measure the ``\textit{days}" variable as the number of days since the last letter (invitation or reminder) instead of the number of days since the invitation letter, because this removes dependency between the ``days" and ``survey phase" variables. For instance, ``Day 2" together with ``survey phase: Invitation" refers to the second day since the expected arrival day of the invitation letter, while ``Day 5" in combination with ``survey phase: Reminder 1" indicates that this is day 5 since receiving the first reminder letter. Together, these two variables specify the baseline hazard rates for all sample cases on a given day.

With ``Invitation" treated as the reference level of \textit{Survey Phase}, the specification of $\gamma_{0s}$ becomes

\begin{equation}
\begin{aligned}
\gamma_{0s} = & \gamma_{00} + \gamma_{01}Days + \gamma_{02}Reminder1 + \\
& \gamma_{03}Reminder2
\end{aligned}
\end{equation}

where $Reminder1 = 1$ if time period $s$ is in the ``Reminder 1" phase and $Reminder1 = 0$ otherwise; likewise, $Reminder2 = 1$ if $s$ is in the ``Reminder 2" phase and $Reminder2 = 0$ otherwise. \textit{Days} remains untransformed, because common transformation of this variable (e.g. log, clog-log, square, cube, square-root) does not lead to better model fit.

\subsection{Lasso Regularisation}
Logistic regression, however, has one shortcoming. It cannot handle the relatively large number of highly correlated predictors in the current data. For instance, there are in the current data 20 weather variables and 10 GT variables, many of whom are highly correlated with each other (e.g. ``\textit{average temperature}", ``\textit{maximum temperature}", ``\textit{disease outbreaks: cold}'', ``\textit{disease outbreaks: influenza}''). Therefore, the inclusion of all the predictors would result in a lack of model parsimony. Both the determination of relevant predictors and the interpretation of parameter estimates become much more difficult. Furthermore, having many predictors may result in an overfit model, because some of the predictors may be capturing noises rather than actual signals. Lastly, multicollinearity can lead to inflated parameter variances and model dependency on the relationship among the highly correlated predictors. 

One solution is a popular machine learning technique called Lasso regularisation. Initially proposed by \textcite{Tibshirani1996}, this technique is capable of performing variable selection (while achieving good prediction) and is also compatible with the generalised linear modelling framework in discrete-time survival analysis \parencite{Tutz2016}. Generally speaking, Lasso regularisation works by adding a penalty term $\lambda$ ($\lambda \geq 0$) to the negative \textit{log-likelihood function $-l$ (Equation 10)}, which has the effect of shrinking parameter estimates towards zero. By doing so, Lasso retains only a small number of important variables (i.e. the ones that have non-zero parameter estimates) in a model and thus results in a more parsimonious and interpretable model. Because this variable selection procedure is automatic, we can also conveniently avoid the use of traditional \textit{p}-values and confidence intervals to judge the relevance of a variable. In addition, by introducing a small bias to the model, Lasso significantly reduces model variance and thereby improves a model's out-of-sample predictive performance. 

The value of $\lambda$ needs to be carefully selected, because up until a certain point, the increase in $\lambda$ is beneficial as it only reduces the variance (and hence avoids overfitting), without losing any important properties in the data. After a certain threshold, however, the model starts losing important properties, giving rise to bias in the model and thus underfitting.

To find the optimal $\lambda$, we followed the advice of \textcite{Hastie2009}, which involves the use of \textit{k}-fold cross-validation (CV) . \textit{k}-fold CV entails randomly dividing the entire set of observations into \textit{k} groups (folds) of approximately equal size. The first fold is treated as a validation set, and the model is fit on the remaining \textit{k}-1 folds. The error measure (e.g. root mean squared error) is then computed on the observations in the held-out fold. This procedure is repeated \textit{k} times: each time, a different fold of observations is treated as a validation set. This process results in \textit{k} estimates of the validation error. Averaging all of these estimates gives the \textit{k}-fold CV error estimate. A typical choice of \textit{k} is 5 or 10, which gives accurate estimates of the validation error while keeping the computation feasible. In this study, we used 10-fold CV. 

Next, we chose a range of $\lambda$ values and computed the 10-fold CV error (i.e. deviance) for each value of $\lambda$. Then, we selected the $\lambda$ value for which the CV error is the lowest. Finally, the model was refitted using all of the available observations and the selected $\lambda$ value. 

There are two further considerations regarding the use of Lasso regularisation. First, the original Lasso algorithm has the disadvantage that its selection of variables can be inconsistent. To solve this problem, \textcite{Zou2006} proposed the \textit{adaptive Lasso}, whose penalty term has the form $\lambda \sum_{j=1}^pw_{j}|\gamma_{j}|$, where $w_{j}$ are weights. He showed that for appropriately chosen data-dependent weights, the adaptive lasso provides consistent variable selection. Following the author's advice, we used Ridge regularisation estimates as weights. Note that Ridge regularisation, similar to Lasso, shrinks parameter estimates. However, unlike Lasso, Ridge regularisation does not reduce any estimate to exactly zero and therefore does not perform variable selection. Second, when using Lasso, one usually assigns a less-than-full-rank dummy-coding procedure to a categorical variable, such that all levels of the variable enter the model as separate variables. This allows Lasso to select what it considers to be appropriate reference levels (i.e. the ones with a zero coefficient). Nevertheless, sometimes Lasso retains all levels of a categorical variable in the model. Without a reference category, the interpretation of the parameter estimates of categorical variables becomes impossible. To avoid this problem, we pre-assigned a reference category to all the categorical variables (see Appendix B) before they entered the model. 

\subsection{Model Evaluation}
As the focus of the study is on the influence of predictors on daily response hazards, it is necessary to evaluate the predictive performance of the models with regards to their prediction of the hazard rates when the models are applied to the test data set. For this purpose, we used \textit{root mean squared error} (RMSE) as the evaluation criterion, which quantifies the distance between the observed and the predicted daily hazards.

Using RMSE, we compared the predictive performance of what we call the ``full model'' (which includes all the time-varying predictors) to that of the ``baseline model'' (which includes only the baseline intercept predictors: ``days'' and ``survey phase''). With the full model where we enter all of the predictors into the model without any interaction effect, we assume that the effects of the predictors do not vary with time. This modelling approach has the advantage that the model is more parsimonious and easier to interpret. However, in reality, the effect of a predictor may depend on time. To account for this possibility, we also built an ``interaction model'', where we include interaction terms between the baseline predictors (``days`` and ``survey phase``) and the time-varying contextual predictors and thereby allow the effects of the model predictors to vary over time. Note that we do not interpret this model in terms of parameter estimates, because the resulting model contains non-zero interaction terms whose corresponding main effects are shrunk to zero. Because of this, the model's parameter estimates become difficult to interpret. However, we can compare still the predictive performance of this interaction model with the other two models.

Furthermore, we plotted the predicted cumulative response rates of the three models against the observed ones.

\subsection{Variable Importance}
In addition to knowing whether the model on the whole predicts well, it is also helpful to know whether a specific predictor predicts well (i.e. so-called ``variable importance"). Specifically, one can evaluate the importance of a variable by calculating the increase in the model’s prediction error after permuting the variable \parencite{Molnar2018}. Permuting the variable breaks the relationship between the variable and the true outcome. Thus, a variable is ``important" if shuffling its values increases the model's prediction error, because in this case the model relies on this variable for better prediction. A variable is ``unimportant" if permuting its values leaves the model error unchanged or worse. 

The permutation algorithm for assessing variable importance we used is based on the work of \textcite{Fisher2018}. For an accurate estimate, we used 20 permutations for each variable and averaged the resulting variable importance scores. We used RMSE as the error measure in calculating variable importance.

\subsection{Software}
We conducted all the analyses in R (version 3.5.0) and R studio (version 1.1.383). We used the package ``glmnet" \parencite{Friedman2010} for the implementation of adaptive Lasso logistic regression.

\end{multicols}

\begin{Figure}
\includegraphics[width=0.9\linewidth]{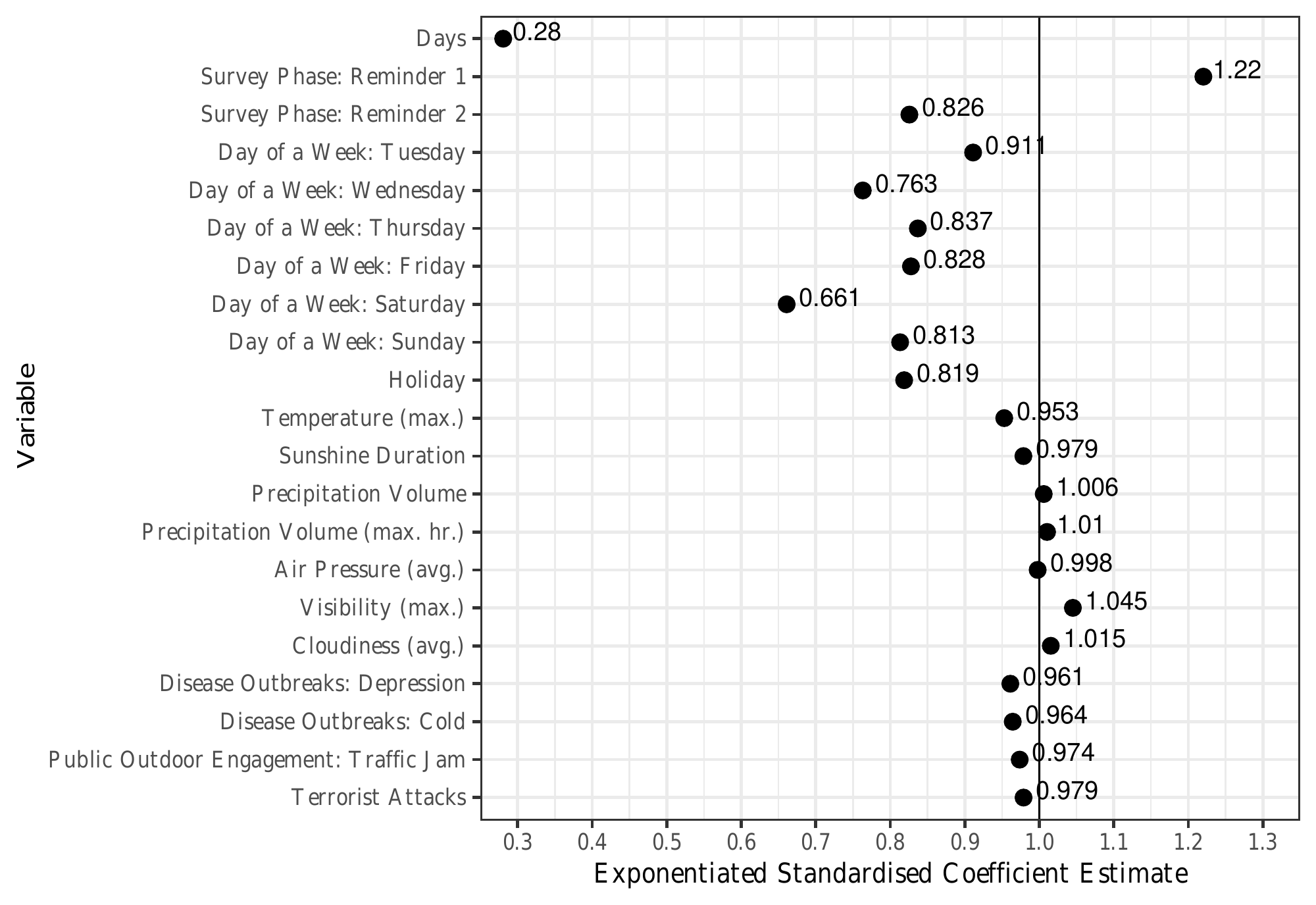}
\centering
\captionof{figure}{Exponentiated Standardised Estimates of Predictors in the Full Model}
\end{Figure}

\begin{multicols} {2}

\section{Results}
\subsection{Model Estimates and Interpretation}
Figure 2 shows the exponentiated standardised parameter estimates of the predictors that are retained by the full model. That is to say, these predictors are considered by the model to have non-zero coefficients and are thus important. The size of the estimates quantifies how much one standard deviation change in the predictors impacts the \textit{conditional} odds of survey response on a given day, under the condition that the person has not responded earlier. 

As the figure suggests, both of the baseline predictors that define the model intercept are strong predictors of survey response. Specifically, the number of \textit{days} since the previous invitation or reminder letter seems to have the largest effect on survey response. An exponentiated standardised coefficient of about 0.28 suggests that, assuming everything else stays constant, for one standard deviation increase in the number of \textit{days} (about 5.34 days), the odds of a person responding to the survey at that given time point are reduced by about 72\%. \textit{Survey phase} also turns out to be an important predictor: the first reminder letter increases the conditional odds of response by about 22\% than in the invitation phase, while the second reminder letter lowers the conditional odds by about 17\%, assuming everything else stays constant. To avoid repetitions, in the rest of the paper we interpret model estimates without repeating the assumption that everything else holds constant.

Turning to the time-varying contextual predictors of interest: \textit{Day of a week} appears as a very relevant predictor. In comparison to Monday, all non-Mondays lower the conditional odds of responses. That is to say, Monday has the most positive effect on conditional response odds, compared to all other days. Saturday shows the strongest negative effect. With an estimate of approximately 0.66, the conditional odds of a survey response on Saturday is about 34\% less likely than on Monday. The effect of Sunday on response odds is also negative, with an exponentiated estimate of 0.81. Therefore, we can safely conclude that weekends have a negative influence on survey response, while Monday has a positive one. 

\textit{Holiday} also appears to have a negative effect on response. With an exponentiated coefficient of 0.82, holidays reduce the conditional response odds by about 18\% compared to non-holidays. 

The weather variables show smaller effects on survey response than the previous variables. Nevertheless, those with non-zero coefficients show a clear pattern. When the weather is nicer (e.g. higher temperature, longer sunshine duration, less rain, higher air pressure, and less cloudy), conditional response odds are also lower. Specifically, for one SD change in these weather variables, conditional response odds can change by a maximum of about 5\%. The only exception to this observed rule is the variable \textit{maximum visibility}, which shows a clear positive effect on response. 

Similar to the weather variables, the GT variables tend to have, if not zero, small effects. Among the variables intended to measure signs of disease outbreaks, ``depression" and ``cold" show negative effects on survey response, while the other indicators of disease outbreaks (``flu``, ``hay fever`` and ``influenza``) are not retained by the model. The two terms concerning data privacy concerns, namely ``data leak" and ``hacking", have also been left out by the model. Between the two variables related to public outdoor engagement, ``traffic jam" negatively predicts survey response, while ``festival`` has a zero coefficient. Finally, ``terrorist" also has a small negative influence on survey response. Note that the interpretation of the GT variables in terms of the sizes of their effects is difficult and can be misleading, because the variables are measured on somewhat arbitrary scales. 

\subsection{Model Performance}
The RMSE scores of the three models (``baseline model'', ``full model'' and ``interaction model'') are 0.005528, 0.005274 and 0.004738, respectively. This suggests that the inclusion of the time-varying contextual predictors in the full model increases the baseline model's predictive performance improves by 4.6\%. In addition, allowing time-varying effects in the interaction model further reduces prediction error by about 10\%, compared with the full model. 

As some may argue that weather variables and/or GT variables largely capture monthly or seasonal trends and therefore can be substituted by indicator variables representing \textit{month} or \textit{season}, we conducted additional analyses to test this argument, which shows that replacing the weather and GT variables with either \textit{month} or \textit{season} indicators lead to poorer performances in RMSE than the models that include the weather and GT variables. Specifically, using month as a replacement variable results in an RMSE score of 0.006553, while using \textit{season} leads to a score of 0.005830. Both are much higher than any of the previous three models we tested. 

Figure 3 presents the predicted cumulative response rates of the three tested models across all survey phases in the test data set, against the observed cumulative response rates (indicated by unconnected asterisks). 

In the invitation survey phase, the interaction model achieves the best prediction of cumulative response rates among the three models. Especially during the later stage of the invitation phase, the interaction model predicts cumulative response rates almost perfectly. Both the full model and the baseline model underpredict cumulative response rates. 

In the first reminder phase, all of the three models seem to predict cumulative response rates better than in the invitation phase. Similar to the invitation phase, the interaction model shows predicted cumulative response rates that are closest to the observed values. The full model and the baseline model have similar prediction performances, with the full model faring slightly better than the baseline model.

In the second reminder phase, the prediction result is more mixed. Around the beginning and the end of the second reminder phase, all of the three models seem capable of predicting cumulative response rates well. However, all the models largely overpredict cumulative response rates in the mid-period, with the interaction model faring slightly better than the other two models. 

Overall, the interaction model achieves the best prediction of cumulative response rates, followed by the full model and then the baseline model. The best prediction is during the first reminder phase. None of the models predict cumulative response rates very well in the second reminder phase.

\end{multicols}

\begin{Figure}
\includegraphics[width=\linewidth]{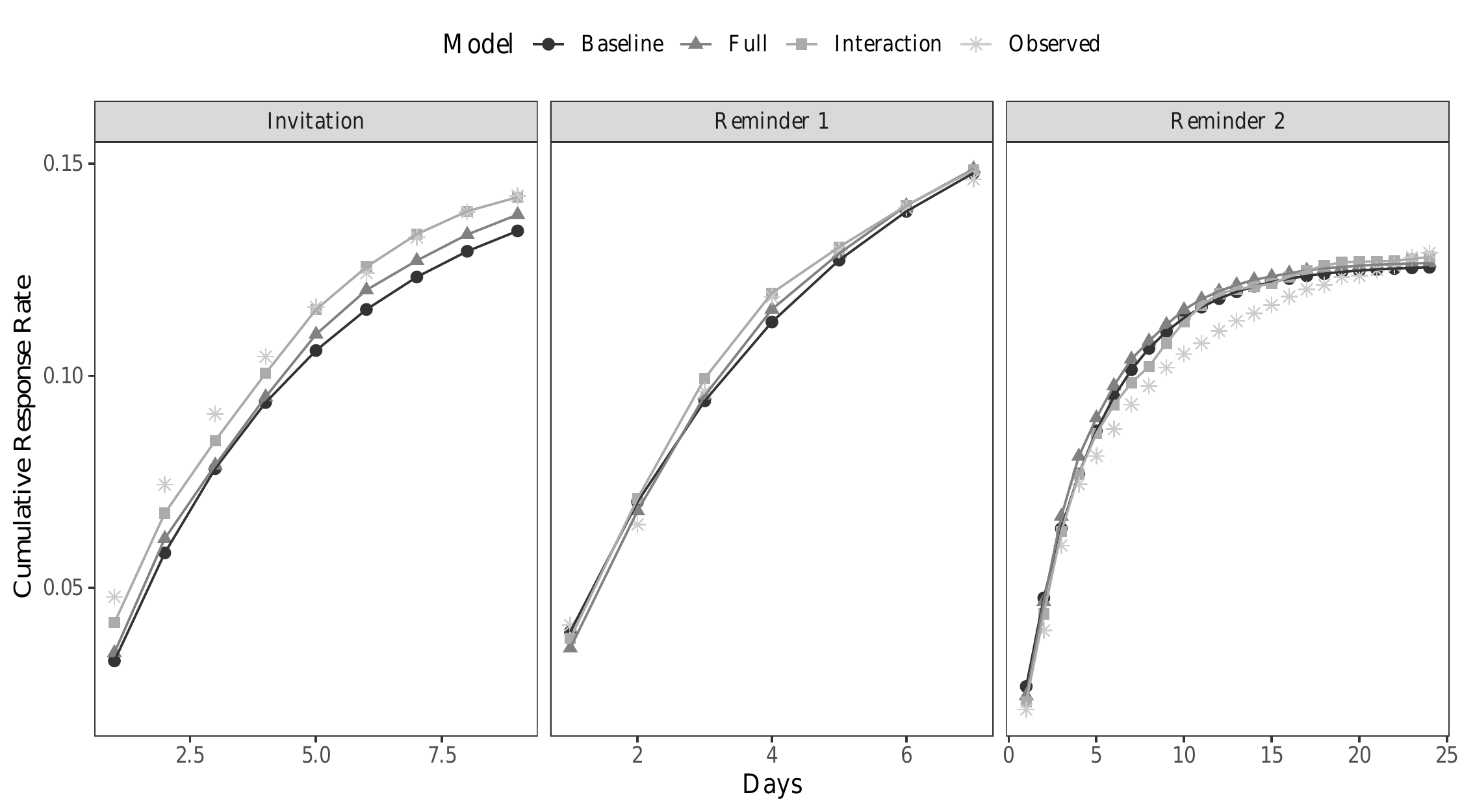}
\centering
\captionof{figure}{Predicted and Observed Cumulative Response Rate by Survey Phase and Model}
\end{Figure}

\begin{multicols}{2}

\subsection{Variable Importance}
Figure 4 shows the variable importance of the predictors with non-zero coefficient estimates from the full model. The variables are displayed in descending order based on their mean variable importance scores. A score above 1 indicates positive contribution to prediction accuracy, while a score below 1 represents the opposite. It is clear that \textit{days} contributes the most to prediction accuracy, followed by the two weekend indicators (i.e. \textit{Saturday} and \textit{Sunday}) and subsequently, the two \textit{survey phase} indicators, namely \textit{reminder 1} and \textit{reminder 2}. The variables representing weekdays, except for Thursday (which has the lowest score among all variables), all contribute to prediction accuracy. Many \textit{weather} variables also contribute to prediction accuracy, especially those measuring temperature, sunshine duration, air pressure, cloudiness and precipitation, despite some of the scores being very small (i.e. only slightly above 1). The \textit{holiday} variable also has a positive contribution; nevertheless, the contribution is very small. The variables below \textit{holidays} have negative variable importance scores. These variables include all of the GT variables, one weather variable (\textit{maximum visibility}) and the \textit{Thursday} indicator.

\end{multicols}

\begin{Figure}
\includegraphics[width=\linewidth]{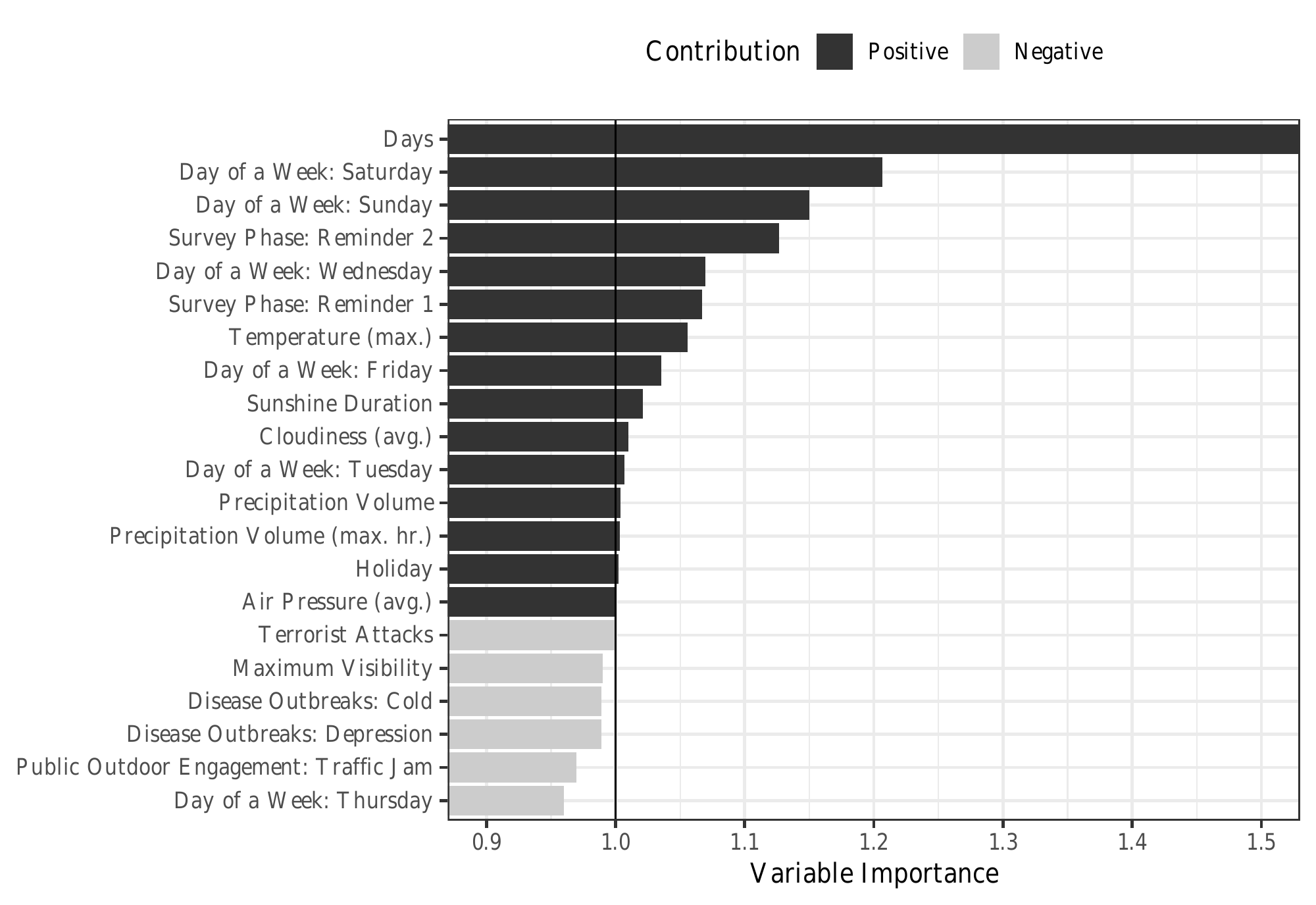}
\centering
\captionof{figure}{Variable Importance of Predictors in the Full Model}
\end{Figure}

\begin{multicols}{2}

\section{Discussion \& Conclusion}
\subsection{Effects of Time-Varying Contextual Predictors}
In this study, we investigate the effects of time-varying contextual factors on web survey response. Specifically, we focus on the explanatory and predictive effects of \textit{day of a week}, \textit{holidays}, \textit{weather}, \textit{disease outbreaks}, \textit{public concerns about data privacy}, \textit{public outdoor engagement} and \textit{terrorism salience} on the response behaviour of the web mode of the Dutch Health Surveys.

Based on our results, \textit{Monday} appears to be the day of a week most positively associated with survey response, followed by Tuesday. All the other days have strong negative effects on response compared to Monday. Among them, Saturday and Sunday show the strongest and the third largest negative effects, respectively, meaning that individuals are less likely to respond during the weekend and would postpone their response until after the weekend, the first day of which is Monday. This may explain why responses are most positively associated with Monday and to a lesser extent, Tuesday. This interpretation is also in line with the previous finding by \textcite{Sauermann2013a} that people would postpone their responses to surveys administered in the weekend until the next week. 

It is also interesting that Saturday has the strongest negative effect on survey response. This might be because on Saturday people have the greatest need among the week to distance themselves away from cognitively demanding activities with low immediate returns so as to properly recover from the work week that has just ended. In addition, people may also have more family and household obligations on Saturday. The work of \textcite{Roeters2018} about time use in 2016 in the Netherlands supports this speculation. The report shows that Saturday is the day among the week when Dutch people spent the most time on household and family care (3.5 hours) and leisure activities (7.6 hours). 

The presence of \textit{holidays} also negatively predicts response decisions. This is in line with the research finding by \textcite{Losch2002} that more contact attempts are required in the holiday season (i.e. summer) for survey research. The underlying mechanism might be similar to the negative effect of Saturday: People spend more time on leisure activities (e.g. travelling and personal care) and family obligations and hence respond less to surveys. 

The weather variables, despite their effects being smaller, also show a clear pattern. When the weather conditions are more pleasant (e.g. higher temperature, longer sunshine duration, less rain, higher air pressure and less cloudy), response odds are lower. This is perhaps due to the common observation that people are more likely to be outdoors when the weather is nice and therefore unable to fill in an online survey that is not smartphone-friendly and that also requires login information from a mailed invitation letter (in the case of the 2016 and 2017 Dutch Health Surveys). The only exception to this observed pattern is the \textit{maximum visibility} variable, which has a positive effect on response that is difficult to interpret. The finding of the negative weather effects on web survey response contrasts the previous finding by \textcite{CunnighamR1979} that more pleasant weather makes people more likely to respond to a face-to-face interview request. Therefore, the influence of weather on survey response is likely moderated by survey modes. 

Among the societal trends variables (i.e. the GT variables), both ``depression'' and ``cold'' negatively predict response decisions, which agrees with our hypothesis that when people are sick, they are less likely to respond. ``Traffic jam", one of the two variables intended to measure public outdoor engagement, is negatively associated with response. This may suggest that on days when many people travel (for work, leisure trips etc.), response rates are lower. This is consistent with the negative effects of good weather and holidays on survey response, because in all three cases the underlying mechanism is likely that people spend more time outdoors and are therefore unlikely to respond. The ``festival'' variable was not retained by the model, which may be because this search term is less correlated with public outdoor engagement than is ``traffic jam". 

The two variables measuring public concerns about data privacy, namely ``data leak'' and ``hacking'', have zero coefficients. This may suggest that in the Netherlands, survey response is not really affected by public concerns over data privacy, or the effect is very little, in contrast to findings from other countries. An alternative explanation might be that these two search terms might not be indicative of the level of public privacy concerns in the Netherlands. 

Terrorism salience also has a negative effect on response, which is in agreement with many research findings that terrorist attacks or threats have many negative consequences for individuals and societies and our hypothesis that it also negatively influences survey response rates. 

Despite being only baseline predictors, the \textit{days} and \textit{survey phase} variables both show interesting and relevant effects on response. Among all the predictors of response in our models, \textit{days} is clearly the strongest one. About five days into a survey data collection process can already reduce a person's response odds by about 70\%. In addition, sending out further reminder letters motivates more survey responses. Interestingly, the positive effect of the first reminder on response odds is stronger than the second reminder (with the invitation letter being the reference group). This suggests that the use of at least one reminder letter may be crucial for a good overall response rate. 

\subsection{Model Prediction}
Overall, all fitted models with time-varying predictors achieve acceptable predictive performances, as indicated by the generally low RMSE scores and their good prediction of cumulative response rates. This suggests that the framework of discrete-time analysis, which is capable of incorporating time-varying effects, is a suitable tool for modelling survey responses and should be recommended for future survey research. 

Among the three models, the interaction model consistently shows the best prediction performance, followed by the full model and lastly, the baseline model. This speaks of, first of all, the relevance of including time-varying predictors like \textit{day of a week}, \textit{holiday} and \textit{weather} for the understanding and the prediction of survey response behaviour. If the goal of the modelling is purely prediction, the approach of the interaction model (i.e. allowing the effects of the predictors to vary over time) will result in better prediction performance, at the cost of model interpretability. Naturally, in this case, even more complex ``black-box'' models like random forests can be utilised. In contrast, if the goal is to obtain an interpretable and parsimonious model, then the approach of the full model without interaction terms may be preferred.

A reduction in RMSE up to 14.2\% (between the baseline and the interaction models) is an incredible outcome. This would allow researchers estimate response rates at the stage of survey planning much more accurately, which can translate into a significant amount of money and resources saved especially for large survey projects.

All of the models moderately overpredict cumulative response rates in the mid-period of the second reminder phase. This requires further examination into, for instance, additional influencing factors of web survey responses and possible nonlinear relationships, for better prediction of responses.

\subsection{Variable Predictive Performance}
The parameter estimates of the predictors were also tested on the test data set, in order to assess variable importance. The results show that \textit{days} is the single most important predictor of survey response. \textit{Survey phase}, \textit{holidays} and the majority of the \textit{day of a week} and \textit{weather} variables also contribute to prediction accuracy with varying degrees. Among the \textit{day of a week} and \textit{weather} variables, two exceptions are the \textit{maximum visibility} and the \textit{Thursday} predictors, which lead to more prediction error. This suggests that \textit{maximum visibility} might be an irrelevant predictor of survey response, at least in our current model. Together with our previous observation that its positive coefficient disagrees with the generally negative effect of good weather on survey response (suggested by the other weather predictor), it might be simply a good idea to leave \textit{maximum visibility} out of the modelling process. The negative variable importance of the latter predictor, \textit{Thursday} (relative to Monday) indicates that Thursday's effect estimate may lack generalisability. However, the \textit{day of a week} variable as a whole (especially the Saturday and Sunday indicators) still contributes strongly and positively to model prediction accuracy, suggesting the overall importance of the \textit{day of a week} variable in prediction modelling of survey response. 

In addition to those variables with negative importance scores, we also notice that some predictors have variable importance scores only slightly above 1, such as \textit{Tuesday}, \textit{precipitation volume (max. hr.)}, \textit{holiday} and \textit{air pressure (avg.)}. There can be several reasons why this is the case. First, it might be because we do not have enough data for the model to obtain coefficient estimates that are more generalisable to unseen data. After all, the trained model is only based on 1.5 years of data from one type of survey. Second, it might be because the Lasso algorithm has introduced too much bias to some model estimates compared to the others. Third, it might be simply that these variables are, indeed, less relevant for prediction of survey responses. In any case, it is important to realise that in any machine learning or statistical models (especially those with many predictors), there are almost always variables with negative variable importance scores because of the imperfection of the algorithm itself, human mistakes during model training, inherent biases and mistakes in the data (e.g. omitted variable bias, multicollinearity) or simply chance. Therefore, we recommend being careful with concluding that certain predictors are definitely and generally irrelevant based on a single study and encourage researching it again in a replication study. 

Lastly, all of the GT variables either do not contribute to prediction accuracy or lead to worse prediction. This is likely due to the fact that GT data are measured on arbitrary scales or that the associated meanings of GT variables may change over time. However, do note that just because the use of the GT variables in this study did not contribute positively to prediction error, it does not necessarily follow that the use of GT variables is always unwarranted or that the found effects of the GT variables are absent. Instead, this only means that GT data may be more suitable for the goal of understanding than prediction, unless a better method (especially in terms of construct validity) for deriving measures from GT data become available.

\subsection{Implication for Survey Data Collection}
In this study, we learn that several time-varying factors may be relevant to web survey response rates. They are: \textit{day of a week}, \textit{holiday}, \textit{weather}, \textit{disease outbreaks} and \textit{public outdoor engagement}. Our results suggest that response rates are likely lower in the following circumstances: during the weekend (especially on Saturday), in the summer (when the weather is generally nice and many people take holidays), during public holidays and lastly, during disease outbreaks (e.g. cold and depression) which might be linked to very cold and dark days in winter. That is to say, heavy data collection during these times might be more difficult. 

Furthermore, we also see from the effect estimates of the two baseline intercept predictors (``days" and ``survey phase") that response odds are highly negatively associated with the number of days since a survey invitation or reminder letter, while they are positively associated with the reminder phases (compared with the invitation phase). This suggests that being able to secure as many responses as possible already in the first few days since the arrival of an invitation or a reminder letter might be crucial for achieving an overall high response rate. Since survey responses, as we have shown, are likely influenced by time-varying contextual factors, it might be beneficial to factor in these potential effects on survey response during the planning phase. In addition, using reminders is also likely to help.

Of course, we should be careful in making causal claims about the found effects of time-varying contextual factors on survey response. Despite reverse causal relationships being impossible here (for instance, whether one responds to a survey or not cannot alter what day it is, the weather or societal trends), nevertheless, this study is only observational in nature (i.e. we did not and cannot vary most of the time-varying contextual factors). We can only tentatively suggest that survey researchers might want to implement their survey projects in times when response rates tend to be higher (e.g. Mondays, workdays, no public holidays, rainy or cloudy days, spring and autumn when the weather is not too nice or too bad). More confidence in such advice can only be warranted by a truly experimental study. 

Lastly, our findings and recommendations may only apply to web surveys whose design, target population and survey culture are similar to the web mode of the 2016 and 2017 Dutch Health Surveys. It is also possible that these findings and recommendations do not hold true for future Dutch Health Surveys, because the survey culture in the Netherlands may as well change. Thus, for survey projects dissimilar to the 2016 and 2017 Dutch Health Surveys, we would recommend that researchers carefully consider the potential influence of time-varying contextual factors on survey response, analyse past survey response data (and if possible, conduct experiments on how survey timing influences response quality) and determine their own data collection practices that are tailored to the needs of their specific project. For surveys similar to the 2016 and 2017 Dutch Health Surveys, we would advise that researchers put our findings and advice to the test of time and more data. 

\subsection{Implication for Survey Methodological Research}
This study employs an innovative idea of using discrete-time survival analysis for modelling survey response processes. One obvious advantage of this approach is its capability of modelling the effects of either time-varying or time-fixed factors alone or both at the same time on an outcome event. This makes modelling of survey response behaviour very flexible, allowing researchers to make use of either kind of variables (or both) available at hand, depending on the availability and accessibility of data. Using only time-varying contextual predictors like the ones in our study has one obvious merit: it does not require sensitive personal information and hence avoids battling with data privacy issues. Our results on model predictive performance further show that modelling with only time-varying contextual predictors can achieve satisfactory prediction of response rates, especially when a model including time-varying (interaction) effects is used. 

In addition, our study employs an interpretable machine learning technique, (adaptive) Lasso regularisation combined with logistic regression, for the purpose of variable selection and good predictive performance. Our results show that the application of Lasso regularisation achieved both purposes. In additional analyses (not shown), we applied gradient boosting (another interpretable machine-learning technique for variable selection and variance reduction) \parencite{Hofner2014} and compared the performance of the resulting gradient boosting models with the performance of the Lasso models. We found that Lasso consistently performed better than the gradient boosting model. Therefore, we recommend the use of (adaptive) Lasso regularisation to survey researchers for the purpose of response modelling.

The study also provides insights into the choice of variables for response modelling. We show that \textit{days}, \textit{survey phase}, \textit{day of a week}, \textit{holiday} and \textit{weather} are important for the prediction of survey responses. In addition, the study makes use of GT data to measure signs of societal trends. While the prediction performance of the GT variables remains questionable, they certainly are convenient tools to capture societal trends of interest that are otherwise difficult to measure. Using GT data thus can help to explore research ideas and explain existing data. 

\subsection{Limitations and Future Direction}
Our study has limitations. For instance, it focuses on the analysis of the web mode of 2016 and 2017 Dutch Health Surveys, thereby making it difficult to generalise the findings to surveys of other modes, in other regions, years, or even other surveys in the Netherlands. The lack of a truly experimental design also undermines the study's internal validity. Therefore, we recommend future research to extend our study framework to more types of variables, surveys, countries and years, ideally with experimental designs. 

We would like to make more suggestions to future research. First, it would be valuable to study how time-varying contextual factors influence different types of non-response behaviour as well, such as no contact and refusals. Second, it would be interesting to study the interaction effects of time-varying contextual factors and respondent-related time-invariant factors on survey response. This would help us to understand, for instance, how someone during school holidays would respond differently to a survey request depending on his or her family status (with or without kids) and occupation (school teacher or not). Third, since each DCP of the Dutch Health Surveys lasts about a month, which is a relatively long period, it would be beneficial to also study how time-varying contextual factors influence much shorter-term surveys. Last but not least, as the current study only focuses on the effects of time-varying factors on survey response rates, future research may want to also examine the corresponding effects on response bias. 

\section{Acknowledgements}
We would like to thank Daniëlle Groffen (Statistics Netherlands) for her help with the used survey data, Dr. Peter Lugtig (Utrecht University) and Dr. Mirjam Moerbeek for their insightful advice to the current research project and Christian Fang MSc (Utrecht University) for his help with proofreading. We have made the data and R code used in this project public at https://doi.org/10.5281/zenodo.4159915. 

\end{multicols}
\printbibliography

\newpage
\begin{centering}
\section*{\hfil Appendix A \hfil}
\end{centering}

\begin{longtable}[c]{lp{13.5cm}}
\caption*{Table: Overview of Evidence of Time-Fixed Factors}
\label{Time-Fixed}\\
\hline
\textbf{Type} & \textbf{Factor} \\
\hline
\endfirsthead
\multicolumn{2}{c}%
{{\bfseries Table \thetable\ continued from previous page}} \\
\hline
\textbf{Type} & \textbf{Factor} \\ \hline
\endhead
Respondent-related & Age, income, household composition (e.g. single, partnered, with children) and ethnic group \parencite{Luiten2013}; education level \parencite{Park2018}; gender and available type of device (e.g. computers, tablets and smartphones) \parencite{KylieBrosnanBettinaGrun2017}; job type and work sector \parencite{Diment2007}; home ownership (own/purchasing, renting) \parencite{Sinclair2012a}; personality traits like conscientiousness, agreeableness and openness to experience \parencite{Marcus2005}; self-perceived expertise to offer relevant information, familiarity with and trust towards survey sponsors, privacy concerns appropriately addressed and propensity for social media participation \parencite{Foster2014}; consecutive participation in surveys (i.e. survey fatigue) \parencite{Porter2004};  \\
\hline
Region-related & Surveyed country, level of population growth, population age structure, internet and cell-phone coverage \parencite{Daikeler2018}; urban density, percentage of foreigners \parencite{Luiten2013}; percentage of rented households, percentage of vacant units, percentage of sing-person households, population density, median house value \parencite{Erdman2016}; \\
\hline
Design-related & Contact mode (mail vs. email), content of subject line, location of URL link, length of the invitation text, and survey time/effort estimate \parencite{Kaplowitz2012}; use of prenotification \parencite{Bandilla2012}; sponsorship \parencite{Fang2012}; personalised invitation, use of pictures in invitation, background colour of invitation \parencite{James2009}; use of reminders \parencite{Goritz2012}; survey length, question difficulty and use of progress bar \parencite{Liu2018}; incentive type and incentive amount \parencite{Coopersmith2018}; survey login method and interval of reminders \parencite{Crawford2001}; percentage of mandatory questions, and the questionnaire item display (i.e. paging or scrolling) \parencite{Tangmanee2019}.\\
\hline
\end{longtable}

\newpage
\begin{centering}
\section*{\hfil Appendix B \hfil}
\end{centering}

\begin{longtable}[c]{m{2.5cm}m{2.5cm}m{11.3cm}}
\caption*{Table: Overview of Time-Varying Variables}
\label{my-label}\\
\hline
\textbf{Type} & \textbf{Variable} & \textbf{Description} \\ \hline
\endfirsthead
\multicolumn{3}{c}%
{{Table: Overview of Time-Varying Variables (Continued)}} \\
\hline
\textbf{Type} & \textbf{Variable} & \textbf{Description} \\ \hline
\endhead
Basic Time & Days & The number of days since the last letter (invitation or reminder) \\ \hline
Basic Time & Survey Phase & The phase in the survey. 3 categories: ``invitation" (reference category), ``reminder 1", ``reminder 2" \\ \hline
Day of a Week & Day of a Week & 7 categories: ``Monday" (reference category), ``Tuesday", ``Wednesday", ``Thursday", ``Friday", ``Saturday", ``Sunday". \\ \hline
Holiday & Holiday & A dummy \{0,1\} indicating whether a day is a public or popular holiday in the Netherlands, which includes New Year's Day, Carnival, Good Friday, Easter, Easter Monday, King's Day, Liberation Day, Ascension Day, Pentecost, Whit Monday, Saint Nicholas's Day, Christmas, and New Year's Eve. \\ \hline
Weather & Temperature & Three variables: maximum, minimum and average temperature (in \degree C) of the day. \\ \hline
Weather & Sunshine Duration & Sunshine duration of the day (in hours) \\ \hline
Weather & Sunshine Percentage & Percentage of sunshine of the day \\ \hline
Weather & Precipitation Volume & Two variables: daily total precipitation volume and hourly maximum precipitation volume (in mm) of the day \\ \hline
Weather & Precipitation Duration & Precipitation duration of the day (in hours) \\ \hline
Weather & Wind Speed & Three variables: maximum, minimum and average hourly wind speed (in m/s) of the day. \\ \hline
Weather & Cloudiness & An ordinal variable, from 1 to 9, indicating the degree of average cloudiness from low to high of the day. \\ \hline
Weather & Visibility & Two variables: minimum and maximum distance of visibility (in m) of the day. \\ \hline
Weather & Humidity & Three variables, measured as maximum, minimum and average humidity (in \%) of the day. \\ \hline
Weather & Air Pressure & Three variables: maximum, minimum and average air pressure (in hPa) of the day. \\ \hline
Google Trends & Disease Outbreaks & Five search terms (with English translation in parentheses): ``depressie" (depression), ``griep (flu)", ``influenza (influenza)", ``verkoudheid (cold)" and ``hooikoorts (hay fever)". \\ \hline
Google Trends & Data Privacy Concerns & Two search terms (with the English translation in parentheses): ``datalek (data leak)" and ``hacking (hacking)". \\ \hline
Google Trends & Outdoor Engagement & Two search terms (with English translation in parentheses): ``files (traffic jam)", ``festival (festival)". \\ \hline
Google Trends & Terrorism Salience & One search term (with English translation in parentheses): ``terrorist (terrorist)". \\ \hline
\end{longtable}

\newpage
\begin{centering}
\section*{\hfil Appendix C \hfil}

\begin{longtable}[c]{m{6.8cm}m{4.9cm}m{4.8cm}}
\caption*{Table: Descriptive Statistics of Variables in the Training and Test Data}
\label{descriptives1}\\
\hline
\textbf{Variable} & \textbf{Training Data} & \textbf{Test Data} \\ \hline
\endfirsthead
\multicolumn{3}{c}%
{{Table: Descriptive Statistics of Variables in the Training and Test Data (Continued)}} \\
\hline
\textbf{Variable} & \textbf{Training Data} & \textbf{Test Data} \\ \hline
\endhead
Days & Mean (SD): 7.33 (5.34) & Mean (SD): 8.26 (5.97) \\ \hline
Survey Phase: Invitation & Count (\%): 138 (23.5) & Count (\%): 54 (24.4) \\ \hline
Survey Phase: Reminder 1 & Count (\%): 125 (21.3) & Count (\%): 42 (19.0) \\ \hline
Survey Phase: Reminder 2 & Count (\%): 324 (55.2)  & Count (\%): 125 (56.6) \\ \hline
Day of a Week: Monday & Count (\%): 86 (14.7) & Count (\%): 30 (13.6) \\ \hline
Day of a Week: Tuesday & Count (\%): 81 (13.8) & Count (\%): 29 (13.1) \\ \hline
Day of a Week: Wednesday & Count (\%): 78 (13.3) & Count (\%): 29 (13.1)  \\ \hline
Day of a Week: Thursday & Count (\%): 81 (13.8) & Count (\%): 34 (15.4) \\ \hline
Day of a Week: Friday & Count (\%): 86 (14.7) & Count (\%): 33 (14.9) \\ \hline
Day of a Week: Saturday & Count (\%): 88 (15.0) & Count (\%): 33 (14.9) \\ \hline
Day of a Week: Sunday & Count (\%): 87 (14.8) & Count (\%): 33 (14.9)  \\ \hline
Holiday & Count (\%): 27 (4.6) & Count (\%): 6 (2.7)  \\ \hline
Temperature (avg.) & Mean (SD): 10.37 (6.24)  &  Mean (SD): 12.10 (5.28)  \\ \hline
Temperature (max.) & Mean (SD): 14.34 (7.09) &  Mean (SD): 15.52 (6.23) \\ \hline
Temperature (min.) & Mean (SD): 6.21 (5.78) &  Mean (SD): 8.49 (4.68)  \\ \hline
Sunshine Duration & Mean (SD): 5.29 (3.77) &  Mean (SD): 3.95 (3.24)  \\ \hline
Sunshine \% & Mean (SD): 41.08 (26.93) &  Mean (SD): 31.60 (23.23) \\ \hline
Precipitation Volume & Mean (SD): 1.60 (2.42) &  Mean (SD): 2.23 (2.65) \\ \hline
Precipitation Volume (max. hr.) & Mean (SD): 0.74 (1.10)  &  Mean (SD): 1.19 (1.38)\\ \hline
Precipitation Duration & Mean (SD): 1.87 (3.06) &  Mean (SD): 3.20 (4.45) \\ \hline
Wind Speed (avg. hr.) & Mean (SD): 4.73 (1.82) &  Mean (SD): 5.01 (2.04)\\ \hline
Wind Speed (max. hr.) & Mean (SD): 7.09 (2.26) &  Mean (SD): 7.48 (2.54)\\ \hline
Wind Speed (min. hr.) & Mean (SD): 2.41 (1.44) &  Mean (SD): 2.62 (1.64) \\ \hline
Cloudiness & Mean (SD): 5.84 (2.01) &  Mean (SD): 6.56 (1.43) \\ \hline
Visibility (max.) & Mean (SD): 75.06 (7.51) &  Mean (SD): 74.81 (7.44)\\ \hline
Visibility (min) & Mean (SD): 39.94 (17.37) &  Mean (SD): 37.92 (16.79) \\ \hline
Humidity (avg.) & Mean (SD): 81.18 (8.34) &  Mean (SD): 84.05 (6.52) \\ \hline
Humidity (max.) & Mean (SD): 94.70 (3.89) &  Mean (SD): 95.62 (2.60) \\ \hline
Humidity (min.) & Mean (SD): 64.67 (13.06) &  Mean (SD): 68.98 (11.42) \\ \hline
Air Pressure (avg.) & Mean (SD): 1016.80 (9.61) &  Mean (SD): 1014.46 (9.00)  \\ \hline
Air Pressure (max.) & Mean (SD): 1019.58 (9.14)  &  Mean (SD): 1017.62 (8.14)\\ \hline
Air Pressure (min.) & Mean (SD): 1013.85 (10.26) &  Mean (SD): 1011.16 (9.95)  \\\hline
Google Trends (``depression") & Mean (SD): 50.43 (14.23)  &  Mean (SD): 46.86 (8.62) \\ \hline
Google Trends (``cold") & Mean (SD):  47.24 (20.73) &  Mean (SD): 49.26 (18.28) \\ \hline
Google Trends (``flu") & Mean (SD): 89.20 (78.88) &  Mean (SD): 59.58 (35.33)\\ \hline
Google Trends (``influenza") & Mean (SD): 24.90 (15.45) &  Mean (SD): 22.12 (17.34) \\ \hline
Google Trends (``hay fever") & Mean (SD): 13.51 (21.84) &  Mean (SD): 2.84 (2.11)\\ \hline
Google Trends (``data leak") & Mean (SD): 19.66 (14.19) &  Mean (SD): 22.66 (16.43)\\ \hline
Google Trends (``hacking") & Mean (SD): 32.48 (8.70) &  Mean (SD): 27.85 (8.82) \\ \hline
Google Trends (``traffic jam") & Mean (SD): 48.07 (14.87) &  Mean (SD): 47.55 (20.11) \\ \hline
Google Trends (``festival") & Mean (SD): 30.58 (16.39)  &  Mean (SD): 27.30 (16.22)\\ \hline
Google Trends (``terrorist") & Mean (SD): 3.95 (4.09) &  Mean (SD): 3.09 (2.38) \\ \hline
\end{longtable}

\end{centering}

\newpage
\begin{centering}
\section*{\hfil Appendix D \hfil}

\begin{longtable}[c]{m{2.5cm}m{3.2cm}m{3.1cm}m{3.2cm}m{3.1cm}}
\caption*{Table: Most Correlated Topics and Queries of the GT Variables}
\label{descriptives2}\\
\hline
\textbf{Search Term} & \textbf{Topic (2016)} & \textbf{Query (2016)} & \textbf{Topic (2017)} & \textbf{Query (2017)}\\ \hline
\endfirsthead
\multicolumn{5}{c}%
{{Table: Most Correlated Topics and Queries of the GT Variables (Continued)}} \\
\hline
\textbf{Search Term} & \textbf{Topic (2016)} & \textbf{Query (2016)} & \textbf{Topic (2017)} & \textbf{Query (2017)}\\ \hline
\endhead
Depression & major depressive disorder - mental disorder & test depressie & major depressive disorder - mental disorder & test depressie \\ \hline
Cold & common cold - disease & tegen verkoudheid & common cold - disease & tegen verkoudheid\\ \hline
Flu & influenza - disease & griep 2016 & influenza - disease & griep 2017\\ \hline
Influenza & influenza - disease & influenza virus & influenza - disease & influenza virus\\ \hline
Hay Fever & hay fever - topic & tegen hooikoorts & hay fever - topic & hooikoorts radar \\ \hline
Data Leak & data breach - topic & datalek melden & data breach - topic & datalek melden\\ \hline
Hacking & hacker - topic & growth hacking & hacker - topic & ethical hacking \\ \hline
Traffic Jam & traffic congestion - topic & file & traffic congestion - topic & file \\ \hline
Festival & festival - topic & festival 2016 & festival - topic & festival 2017\\ \hline
Terrorist & terrorism - type of criminal organisation & terrorist attack & terrorist attack - disaster type & terrorist attack \\ \hline
\end{longtable}
\end{centering}

Translation of Dutch Terms to English: test depressie (depression test), tegen verkoudheid (against cold), griep (flu), tegen hooikoorts (against hay fever), datalek melden (report data leaks). 

Difference between ``Topic" and ``Query": ``Topic" refers to topics that other users using the same search term also look for; ``Query" refers to search terms that users using the same search term also search for.

\newpage
\begin{centering}
\section*{\hfil Appendix E \hfil}
\end{centering}
As the GT's server allows only a maximum of 244 daily scores per inquiry per search term, we obtained, for every search term, 244 observations from 244 samples for each date in 2016 and 2017, with each sample shifted by a single day relative to the previous one. With 244 observations per date, we used in total 974 samples to cover all the dates in 2016 and 2017. The first sample covers the dates between ``2015-05-03" and ``2016-01-01" and the last between ``2017-12-31" and ``2018-08-31". 
We introduce the following notation for one GT query: $GT_s := \{G_{s,i}, i \in \{1:244\} \}$ for sample $s$; $G_{s1}$ is the index score of the first day in the requested period and $G_{s244}$ the last day. The index $s$ runs from 1 to 974 with $s = 1$ for the sample that starts on $t_1 :=$ 2015-05-03 and $s = 974$ for the sample that starts on $t_{974} :=$ 2017-12-31. The first day of interest is $t_{244} := $ 2016-01-01. By writing out all samples below each other, it is easy to see that the following index scores belong to day $t_{244}$: $GT(t_{244}) = \{ G_{1,244}, G_{2,243}, ..., G_{244,1} \}$. In general, the following set of index scores belongs to day $t_n$: $GT(t_n) = \{ G_{s,i}, s \in \{1:974\}, i \in \{1:244\} | (s + i = n + 1) \}$. For each day of interest ($n \in \{244:974\}$) the set size $m(t_n)$ of relevant GT index scores for that day equals 244. An averaged GT index score for a day $t_n$, $\overline{P}(t_n)$ is then calculated by averaging the calibrated values in $GT(t_n)$, where the calibration factors $w_s$ for an index score from sample $s$ still has to be determined. This leads to the following formula:

\begin{equation} \label{eq1}
\begin{split}
\overline{P}(t_n) & = \frac{1}{244} \sum_{(s,i) | (s+i = n+1)} G_{s,i} \cdot w_s \\
                  & = \frac{1}{244} \sum_{s=1}^{974} \sum_{i=1}^{244} G_{s,i} \cdot w_s \cdot \delta_{(s+i),(n+1)}
\end{split}
\end{equation}

The calibration factor is determined by looking at the overlap between two consecutive GT samples. There is an overlap of 243 days and the sum of all index scores of those overlapping days is compared. The first GT query $GT_1$ is taken as a reference. The second series is then calibrated to the first series by a factor $C_2$ for which:
\begin{equation}
\sum_{i=2}^{244} G_{1,i} = C_2 \sum_{j=1}^{243} G_{2,j} \Rightarrow C_2 := \frac{ \sum_{i=2}^{244} G_{1,i} }{ \sum_{j=1}^{243} G_{2,j} }
\end{equation}
In general, $C_k$ is determined by the following equation for $k \geq 2$:
\begin{equation}
C_k := \frac{ \sum_{i=2}^{244} G_{k-1,i} }{ \sum_{j=1}^{243} G_{k,j} }
\end{equation}
For $k=1$ define $C_1 := 1$.

By induction one can then show that each query $GT_s$ is calibrated to the first query $GT_1$ by the calibration factor:
\begin{equation}
w_s = \prod_{k=1}^{s} C_k .
\end{equation}
In total this gives:
\begin{equation}
\overline{P}(t_n) = \frac{1}{244} \sum_{s=1}^{974} \sum_{i=1}^{244} G_{s,i} \cdot  \prod_{k=1}^{s} C_k \cdot \delta_{(s+i),(n+1)}
\end{equation}
with $C_k$ as defined above.

\end{document}